\documentclass[entropy,article,accept,pdftex,moreauthors]{Definitions/mdpi} 
\usepackage{amsmath}
\usepackage{ulem}
\firstpage{1} 
\pubvolume{1}
\issuenum{1}
\articlenumber{0}
\pubyear{2025}
\copyrightyear{2025}
\datereceived{ } 
\daterevised{ }
\dateaccepted{ } 
\datepublished{ } 
\hreflink{https://doi.org/}

\pdfoutput=1

\Title{Detrended cross-correlations and their random matrix limit: an example from the cryptocurrency market}

\TitleCitation{Detrended cross-correlations and their random matrix limit}

\Author{Stanisław Drożdż $^{1,2,}$*\orcidA{}, Paweł Jarosz $^{2}$\orcidB{}, Jarosław Kwapień $^{1}$\orcidC{}, Maria Skupień $^{3}$\orcidD{}, Marcin Wątorek $^{2}$\orcidE{}}

\AuthorNames{Stanisław Drożdż, Paweł Jarosz, Jarosław Kwapień, Maria Skupień and Marcin Wątorek}

\address{%
$^{1}$ \quad Complex Systems Theory Department, Institute of Nuclear Physics, Polish Academy of Sciences, ul. Radzikowskiego 152, 31-342 Kraków, Poland;\\
$^{2}$ \quad Faculty of Computer Science and Mathematics, Cracow University of Technology, ul. Warszawska 24, 31-155 Kraków, Poland;\\
$^{3}$ \quad Department of Mathematics, University of the National Education Commission, Podchorążych 2, 30-084, Kraków, Poland;}

\corres{Correspondence: marcin.watorek@pk.edu.pl; stanislaw.drozdz@ifj.edu.pl;}

\abstract{Correlations in complex systems are often obscured by nonstationarity, long-range memory, and heavy-tailed fluctuations, which limit the usefulness of traditional covariance-based analyses. To address these challenges, we construct scale and fluctuation-dependent correlation matrices using the multifractal detrended cross-correlation coefficient $\rho_r$ that selectively emphasizes fluctuations of different amplitudes. We examine the spectral properties of these detrended correlation matrices and compare them to the spectral properties of the matrices calculated in the same way from synthetic Gaussian and $q$Gaussian signals. Our results show that detrending, heavy tails, and the fluctuation-order parameter $r$ jointly produce spectra, which substantially depart from the random case even under absence of cross-correlations in time series. Applying this framework to one-minute returns of 140 major cryptocurrencies from 2021–2024 reveals robust collective modes, including a dominant market factor and several sectoral components whose strength depends on the analyzed scale and fluctuation order. After filtering out the market mode, the empirical eigenvalue bulk aligns closely with the limit of random detrended cross-correlations, enabling clear identification of structurally significant outliers. Overall, the study provides a refined spectral baseline for detrended cross-correlations and offers a promising tool for distinguishing genuine interdependencies from noise in complex, nonstationary, heavy-tailed systems.}

\keyword{Multifractal cross-correlations; Detrended cross-correlation analysis; random matrix theory; eigenvalue spectra; cryptocurrency market} 

\begin{document}

\section{Introduction}

Correlations among complex systems’ components play a central role in understanding their collective dynamics~\cite{KwapienJ-2012a}. In fields ranging from finance~\cite{BunJ-2017,JiangZQ-2019a,BardosciaM-2021a}, climatology~\cite{SrikanthanR-2001a,AlKandariNM-2005a,FeijooA-2016a,MengisN-2018}, molecular and biological systems~\cite{DykemanEC-2010a,KamberajH-2011a} to neuroscience~\cite{KwapienJ-2000,DeChaveigneA-2019a,SnyderA-2022}, and physics~\cite{IzenmanAJ-2021}, correlation matrices serve as key tools for quantifying interdependencies between multivariate time series. However, the presence of nonstationarities and long-range dependencies in empirical data often leads to spurious correlations, challenging traditional covariance-based analyses~\cite{PengCK-1994a,ChenZ-2002,BryceRM-2012a}. To address these issues, detrended cross-correlation analysis (DCCA)~\cite{PodobnikB-2008a} and its generalizations~\cite{ZhouWX-2008a,OswiecimkaP-2014a,ZebendeGF-2011a} have been developed as robust methods for quantifying power-law cross-correlations between nonstationary signals.

Correlation analysis-based methods are widely used in financial markets. They have been successfully applied to stock markets~\cite{Drozdz2000,Plerou2000,PlerouV-2002a,Utsugi2004,Wang2013,ZhaoL-2018a,James2021chaos,James2022PhysAstock}, forex~\cite{drozdz2007,Mai2018,GebarowskiR-2019a,Miskiewicz2021,MiskiewiczJ-2022a}, cryptocurrencies~\cite{StosicD-2018a,Basnarkov2019,ZiebaD-2019a,DrozdzS-2020b,BriolaA-2022a,JamesN-2022a,JamesN-2022b,JingR-2023a,Watorek2023,Jin2025}, and even NFT tokens~\cite{WatorekM-2024a}. These methods are useful in trading, risk management and portfolio optimization~\cite{CohenG-2022a,JamesN-2022a,JamesN-2022b,JamesN-2022s,James2023ec,Bhattacherjee2025,sila2025,Tsioutsios2025}.

In this work, we study the spectral properties of matrices constructed from detrended cross-correlation coefficients $\rho_r$ (also denoted in literature as $\rho_q)$, which generalize the detrended cross-correlation coefficient $\rho_{_{\rm DCCA}}$~\cite{ZebendeGF-2011a}, which is an equivalent of the Pearson cross-correlation for the detrended fluctuation analysis~\cite{KwapienJ-2015a}. Each element of the resulting matrix captures the scale-dependent correlation between detrended fluctuations of two time series, controlled by the parameter $r$ that governs sensitivity to fluctuation amplitudes. This approach allows us to explore interdependencies in systems characterized by heterogeneous scaling behaviors or multifractality, going beyond the scope of linear correlation measures.

Our primary interest lies in the eigenvalue spectra of such detrended correlation matrices. In conventional correlation matrix theory, the Marčenko–Pastur (M-P) distribution provides the null hypothesis for the eigenvalue density when entries are independent and identically distributed random variables, i.e. when no genuine correlations are present~\cite{MarcenkoVA-1967a}. However, when matrix elements are derived from detrended cross-correlations, the statistical structure may differ substantially from that assumed by the M-P framework. The detrending procedure introduces scale-dependent filtering and residual cross-structure even under nominally uncorrelated conditions. Consequently, the classical random matrix limit cannot be directly applied as a baseline for identifying significant correlations.

We therefore investigate the eigenvalue spectra of the detrended correlation matrices, constructed from the ensemble of synthetic time series representing Gaussian and $q$Gaussian models~\cite{UmarovS-2008a}. We aim to characterize how detrending and fluctuation-based weighting modify the spectral density and the onset of collective modes. The results provide insights into the appropriate null hypotheses for spectral analysis of detrended correlations and offer a refined framework for distinguishing genuine collective behavior from stochastic background fluctuations in complex systems.

As a practical application of this formalism, we analyze a representative set of the most liquid 140 cryptocurrencies from Binance exchange~\citep{Binance}. The cryptocurrency market provides a particularly suitable testing ground for detrended correlation analysis due to its pronounced nonstationarity, strong cross-dependencies, and heterogeneity in trading activity and capitalization~\cite{WatorekM-2021b}. Price series of digital assets exhibit complex temporal structures, including volatility clustering~\cite{DrozdzS-2018a,JamesN-2021a,EvrimMandaciP-2022a,NguyenAPN-2023a,Brouty2024,Sila2024,Queiroz2024,BuiHQ-2025a} and multifractal scaling~\cite{TakaishiT-2018a,Stosic2019,TakaishiT-2020a,Bariviera2021,KwapienJ-2022a,KwapienJ-2022b,Watorek2024,Drozdz2025}, which render standard correlation measures inadequate. Their price changes are also susceptible to external influences from other financial markets~\cite{ConlonT-2020a,JamesN-2021a,ZhangYJ-2021a,ChoiS-2022a,ElmelkiA-2022a,WatorekM-2023a,Kristjanpoller2025,Li2025,NguyenAPN-2025a}, as well as geopolitical shocks~\cite{BouriE-2022a,HongMY-2022a,KhalfaouiR-2023a,FangY-2024a}  or social media impact~\citep{PoongodiM-2021a,AharonDY-2022a}.

By constructing the detrended correlation matrix $\boldsymbol{\rho}_r(s)$ from the $\rho_r(s)$ coefficients computed across multiple time scales $s$, we are able to probe the underlying architecture of interdependencies among these assets while suppressing the influence of global trends and nonstationary effects. Examination of the eigenvalue spectra reveals the presence of collective modes associated with market-wide behavior, sectoral groupings, and noise-dominated components. Comparing these empirical spectra with the corresponding limits derived for uncorrelated signals allows us to identify statistically significant deviations indicative of genuine market structure. This approach thus offers a refined spectral perspective on the collective dynamics and information flow within the cryptocurrency market.

\section{Methods}

\subsection{Detrended cross-correlations}

We start from a set of time series: ${\rm U}_i = \{u_i(j)\}_{j=1}^T$ with $i=1,...,N$. For each time series, a corresponding signal profile is constructed by integrating the data along the time axis:
\begin{equation}
\bar{u}_i(k) = \sum_{j=1}^k u_i(j), \quad k=1,...,T.
\label{eq::integration}
\end{equation}
Next, for a given scale $s$, this signal profile is divided into $2 M_s$ non-overlapping windows of width $s$ starting from both ends of the time series, and, subsequently, it is detrended by best-fitting a polynomial $P_{\nu}^{(m)}$ of order $m$ in each window $\nu$ individually:
\begin{equation}
x_i(s\nu+k) = \bar{u}_i(s\nu+k)-P_{\nu}^{(m)}(k), \quad k=1,...,s, \quad \nu=0,...,2M_s-1.
\label{eq::detrending}
\end{equation}
By proceeding along this way with all time series, we obtain a set of the detrended time series ${\rm X}_i$, where $i=1,...,N$ that will be subject to further analysis. Signal covariance is then calculated in each window:
\begin{equation}
f^2_{ij}(s,\nu) = \frac{1}{s} \sum_{k=1}^s x_i(s\nu+k) x_j(s\nu+k), \quad i,j=1,...,N,
\label{eq::covariance}
\end{equation}
which becomes variance if $i=j$. In order to allow for a multiscale analysis, a parameter $r \in \mathbb{R}$ is introduced and a family of bivariate $r$-fluctuation functions is introduced:
\begin{equation}
F^r_{ij}(s) = \frac{1}{2 M_s} \sum_{\nu=0}^{2M_s-1} {\rm sign}[f^2_{ij}(s,\nu)] |f^2_{ij}(s,\nu)|^{r/2},
\label{eq::bivariate.fluctuation.function}
\end{equation}
where sign of each covariance is preserved while the modulus allows us to avoid complex values if $f^2_{ij}(s,\nu) < 0$ and $r/2$ is not integer~\cite{OswiecimkaP-2013a}. These functions become univariate if $i=j$. Based on Eq.~(\ref{eq::bivariate.fluctuation.function}), it is possible to define an $r$-dependent detrended cross-correlation coefficient~\cite{KwapienJ-2015a}:
\begin{equation}
\rho_r^{ij}(s) = \frac{F_{ij}^r(s)}{\sqrt{F_{ii}^r(s) F_{jj}^r(s)}},
\label{eq::rho.r}
\end{equation}
whose values $-1 \leqslant \rho_r^{ij}(s) \leqslant 1$ have similar interpretation as the Pearson cross-correlation coefficient in the case of $r>0$. However, by considering different values of $r$, we are able to focus on the cross-correlations among specific range of fluctuation amplitudes: large-amplitude fluctuation contribution is amplified if $r > 2$ and small-amplitude fluctuation contribution is amplified if $r < 2$. 

For a given set of time series, the coefficient $\rho_r^{ij}(s)$ is symmetric with respect to time series order and it can be calculated for all pairs $(i,j)$ with $j=i+1,...,N$. An $r$-dependent detrended cross-correlation matrix is then constructed as
\begin{equation}
\boldsymbol{\rho}_r(s) = [\rho_r^{ij}(s)]_{i,j=1}^N.
\label{eq::rho.r.matrix}
\end{equation}
The whole procedure from the detrending step to the matrix construction can be repeated for different scales $s$ in some range $s_{\rm min} \leqslant s \leqslant s_{\rm max}$.

\subsection{Spectral characteristics of the detrended cross-correlation matrix}

Spectral properties of $\boldsymbol{\rho}_r$ can be determined by solving the eigenvalue problem:
\begin{equation}
\boldsymbol{\rho}_r(s) {\bf v}_i(s) = \lambda_i(s) {\bf v}_i(s), \quad i=1,...,N,
\end{equation}
where the ordering is such that $\lambda_1 \geqslant \lambda_2 \geqslant ... \geqslant \lambda_N$. While the eigenvalues of $\boldsymbol{\rho}_r$ are real-valued owing to the symmetry of the matrix, this matrix is not necessarily positive semi-definite. First, let us look at the simplest case of $r=2$ and construct a matrix of the fluctuation functions (\ref{eq::bivariate.fluctuation.function}) for all pairs of the time series:
\begin{equation}
{\bf F}^2(s) = \frac{1}{M_s} \sum_{\nu} {\bf f}^2(s,\nu) = \frac{1}{M_s} \sum_{\nu} \frac{1}{s} {\bf X}_{\nu}^{\rm T}(s) {\bf X}_{\nu}(s) = \frac{1}{T} {\bf X}^{\rm T}(s) {\bf X}(s),
\end{equation}
where ${\bf X}_{\nu}$ is a data matrix created from all time series in a window $\nu$ and ${\bf X}$ is an analogous data matrix for the whole time series (to simplify notation, we assume that their length $T$ is a multiple of $s$). A dependence of ${\bf X}$ on scale $s$ comes from the fact that detrending a signal with a polynomial of a given degree leaves the more trend in the residual signal the larger the scale $s$ is. Since it can be expressed as the product ${\bf X}^{\rm T} {\bf X}$, the matrix ${\bf F}^2$ is positive semi-definite. 

Now, let us consider the case of $r=2n$, $n \in \mathbb{N}$, in which
\begin{equation}
{\bf F}^{2n}(s) = \frac{1}{M_s} \sum_{\nu} \big[ {\bf f}^2(s,\nu)\big]^{\circ n} = \frac{1}{M_s} \sum_{\nu} \frac{1}{s^n} \big[ {\bf X}_{\nu}^{\rm T}(s) {\bf X}_{\nu}(s) \big]^{\circ n},
\end{equation}
where $[\cdot]^{\circ n}$ denotes the Hadamard product of $n$ matrices $\cdot$ (i.e., their element-wise multiplication). It can be shown that this operation preserves the property of positive semi-definiteness (the Schur theorem~\cite{JohnsonCR-1974a}). A sum of matrices preserves the positive semidefiniteness of the components, therefore there exists a matrix ${\bf \widetilde{X}}(s)$ such that
\begin{equation}
{\bf F}^{2n}(s) = \frac{1}{s^n M_s} \widetilde{\bf X}^{\rm T}(s) \widetilde{\bf X}(s).
\end{equation}
This is generally not true for $r \neq 2n$, however. If this is the case, a condition $r/2 \geqslant N-2$ must hold to guarantee positive semi-definiteness of ${\bf F}^r(s)$ as stated by the FitzGerald-Horn theorem~\cite{FitzGeraldCH-1977a}. Otherwise, this matrix may have negative eigenvalues. When they are negative indeed and when the whole spectrum is non-negative is a delicate question: typically, the smaller is $r$ (especially if $r < 1$), the larger is the probability that some eigenvalues fall below 0. The opposite is true if $r$ increases and, eventually, the matrix enters the $r/2 \geqslant N-2$ region. Finally, based on Eq.~(\ref{eq::rho.r}), we arrive at the conclusion that $\boldsymbol{\rho}_r(s)$ is positive semi-definite if ${\bf F}^r(s)$ reveals this property.

\subsection{Random matrix limit and departures}

The property of positive semi-definiteness of $\boldsymbol{\rho}_r(s)$ is important from the statistical point of view, because we may then ask a question how is a given matrix related to the Wishart ensemble of random correlation matrices ${\bf W}$~\cite{WishartJ-1928a}. In the absence of genuine cross-correlations in ${\bf X}$, this relation is straightforward for $r=2$, because the structure of ${\bf F}^2(s)$ is in this situation similar to the structure of a Pearson correlation matrix. However, for $r \neq 2$ the matrix structure of $\boldsymbol{\rho}_r(s)$ requires applying of different random matrix ensembles.

If the Wishart case is valid, there is a well-defined distribution of eigenvalues of a matrix constructed for a set of uncorrelated Gaussian-distributed time series in the thermodynamical limit $N,T \to \infty$, $T/N = Q = {\rm const}$ (the Marčenko-Pastur distribution~\cite{MarcenkoVA-1967a}):
\begin{equation}
\phi_W(\lambda) = \frac{1}{2 \pi Q \lambda} \sqrt{(\lambda_{\rm max} - \lambda)(\lambda - \lambda_{\rm min})}, \quad \lambda_{\rm min} \leqslant \lambda \leqslant \lambda_{\rm max},
\label{eq::marcenko-pastur}
\end{equation}
where $\lambda_{\rm min}=(1-\sqrt{Q})^2$ and $\lambda_{\rm max}=(1+\sqrt{Q})^2$.

In the case of detrended correlations considered here, however, some deviations from the M-P distribution may arise even for random series due to detrending of fluctuations~\cite{ChenZ-2002}. The space for such deviations increases with increasing the range of detrending (longer scale $s$). To obtain appropriate null hypotheses for the empirical correlations under consideration, in the absence of analytical formulas for the eigenvalue distributions of correlation matrices constructed in such a generalized manner, numerical simulations are necessary. Therefore, such simulated eigenvalue distributions will be conducted in parallel to such analyses for the empirical data.

Another factor that, in the presence of detrending, may lead to further modification of the above characteristics is the fluctuation distributions of the analyzed time series. Distributions of fluctuations of time series representing many natural phenomena develop thicker tails than the ones obeying the normal distribution. For instance, the distributions of fluctuations in financial rates of return on short time scales quite universally~\cite{WatorekM-2021a} satisfy the so-called inverse-cubic power law~\cite{GopikrishnanP-1998a} and this applies even to the cryptocurrency markets~\cite{StosicD-2018a,JamesN-2021b,WatorekM-2023b}. Such distributions are quite efficiently described by functions called $q$Gaussians~\cite{TsallisC-2009a}.

\subsection{$q$Gaussian distribution}

In general $q$Gaussian distributions provide a highly practical analytical framework for modeling the corresponding probability distributions. They constitute a natural extension of the standard Gaussian distribution in analogy of how the Tsallis entropy $S_q$ extends the classical Boltzmann-Gibbs entropy $S$~\cite{UmarovS-2008a,TsallisC-1995a}. This class of distributions $\mathcal{G}q$ is characterized by two parameters: a shape parameter $q \in (-\infty, 3)$ and a scale (width) parameter $\beta > 0$. The corresponding probability density function (PDF) assumes the form~\cite{UmarovS-2008a}:
\begin{equation}
\phi_q(x) = \frac{\sqrt{\beta}}{C_q} e_q(-\beta x^2),
\label{eq::qgaussian.pdf}
\end{equation}
where $e_q(x)$ denotes the $q$-exponential function
\begin{equation}
e_q(x) = \begin{cases}
(1+(1-q)x)^{1/(1-q)} & \text{if } q\neq 1 \text{ and } 1+(1-q)x>0\\
0 & \text{if } q\neq 1 \text{ and } 1+(1-q)x\leq 0\\
e^x & \text{if } q=1,
\end{cases}
\end{equation}
and $C_q = \int{-\infty}^{\infty} e_q(x^2),dx$ is a normalization constant.

Asymptotic behavior at large $|x|$ of a $q$Gaussian for $q=3/2$ then corresponds to the inverse-cubic power law~\cite{DrozdzS-2010a}.  

\subsection{Outlying eigenvalues and collectivity}

The above considerations concern a set of independently generated random time series, within and between which there are no synchronous correlations. However, the essence of real, natural complex systems is the coexistence of randomness with collective synchronous effects. For fundamental and practical reasons, the latter are, of course, the primary object of interest. Within the general formalism of correlation matrices, such effects manifest themselves in significant deviations from the spectrum of eigenvalues assumed by the appropriate null hypothesis as indicated above.  
More specifically~\cite{DrozdzS-2001,DrozdzS-2002,Guhr_2003}, when coherent or collective behavior arises, it effectively reduces the rank of the correlation matrix: a small number of dominant modes capture most of the system’s variance, while the remaining degrees of freedom behave as random noise. This reduced rank is reflected in the presence of a few large eigenvalues that stand well above the random matrix bounds, separating them from the random bulk of the spectrum. Such spectral segregation reveals the emergence of collective dynamics and the formation of correlated groups or patterns within the system. In this way, deviations from full-rank randomness — i.e., the appearance of a low effective rank — serve as a quantitative signature of collective effects and organized structure emerging from an otherwise random background.

\section{Data}

The empirical dataset consists of $N=140$ exchange rates for the most actively traded cryptocurrencies, quoted in USDT on the Binance exchange~\citep{Binance}. It covers the period from January 1, 2021, to September 30, 2024, with data publicly available in an open repository~\citep{DataBinance}. The time series were recorded at a 1 min sampling frequency and transformed into logarithmic returns $R(t_k) = \ln p(t_{k+1}) - \ln p(t_k)$, where $k=1,...,T$.

\begin{figure}[ht!]
\centering
\includegraphics[width=0.99\textwidth]{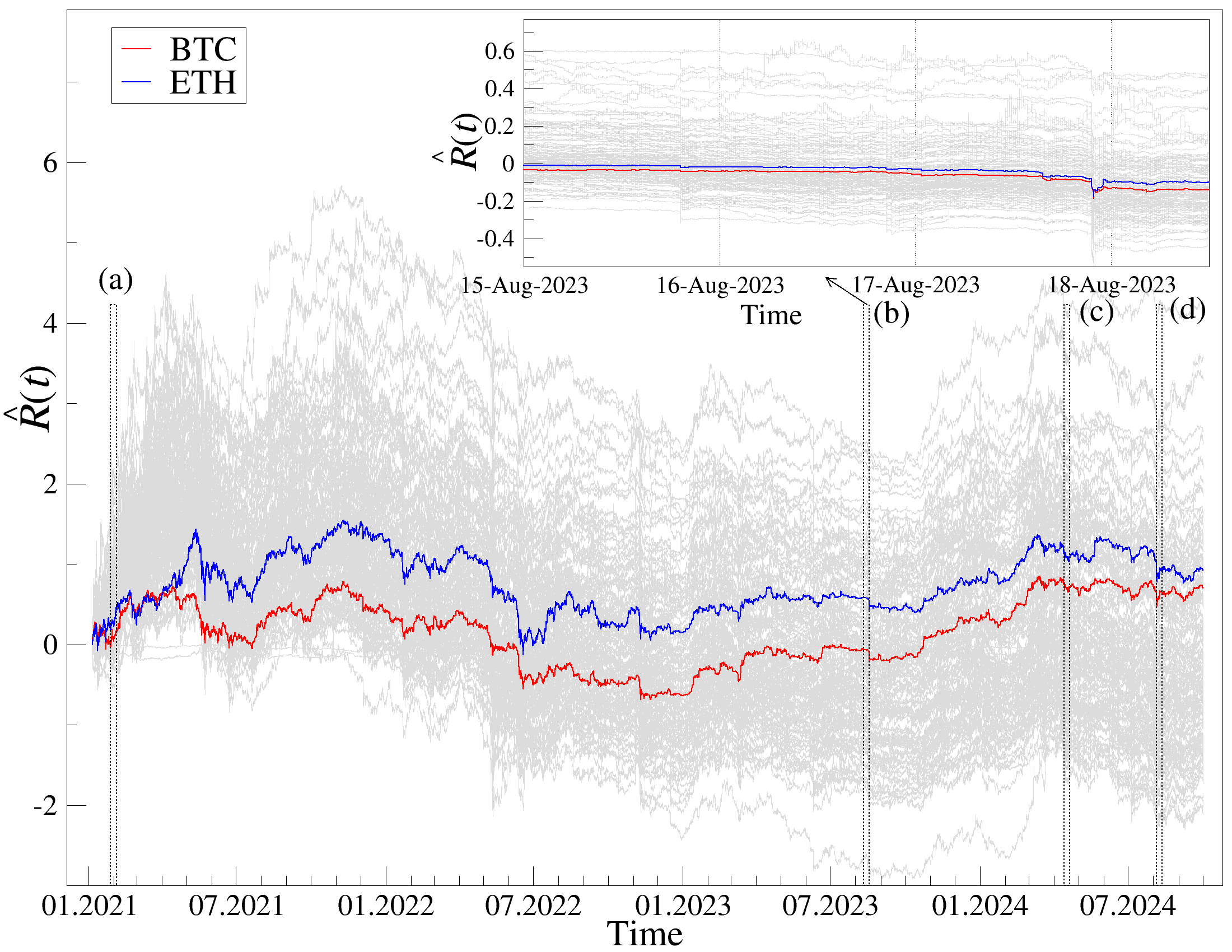}
\caption{(Main) Evolution of the cumulative log-returns $\hat{R}_i(t_k)$ of the 140 cryptocurrencies over the whole time period from Jan 1, 2021 to Sep 30, 2024. The bulk of the cryptocurrencies is shown in the background (grey lines), while BTC and ETH are distinguished by red and blue lines, respectively. Four specific periods are distinguished by narrow vertical rectangles (a)-(d). (Inset) Evolution of the same data during a selected shorter period (b): Aug 15, 2023 to Aug 18, 2023.}
\label{fig::returns.integrated}
\end{figure}

In order for the relative price changes to be directly comparable, the evolution of the cumulative logarithmic returns $\hat{R}_i(t_k)=\sum_{k=1}^T R_i(t_k)$ of the $N=140$ cryptocurrencies considered here indexed by $i$ ($i=1,...,N$) over the analyzed time period is presented in Fig.~\ref{fig::returns.integrated}. Various phases of the market can be observed. The bull market in 2021, then the bear market in 2022 with the crash in May 2022. After reaching its bottom at the end of 2022, the market was in a slower growth phase until mid-2024, then moved sideways until the end of September 2024. Thus, the selected dataset allows for market analysis under various conditions.


\begin{figure}[ht!]
\includegraphics[width=0.99\textwidth]{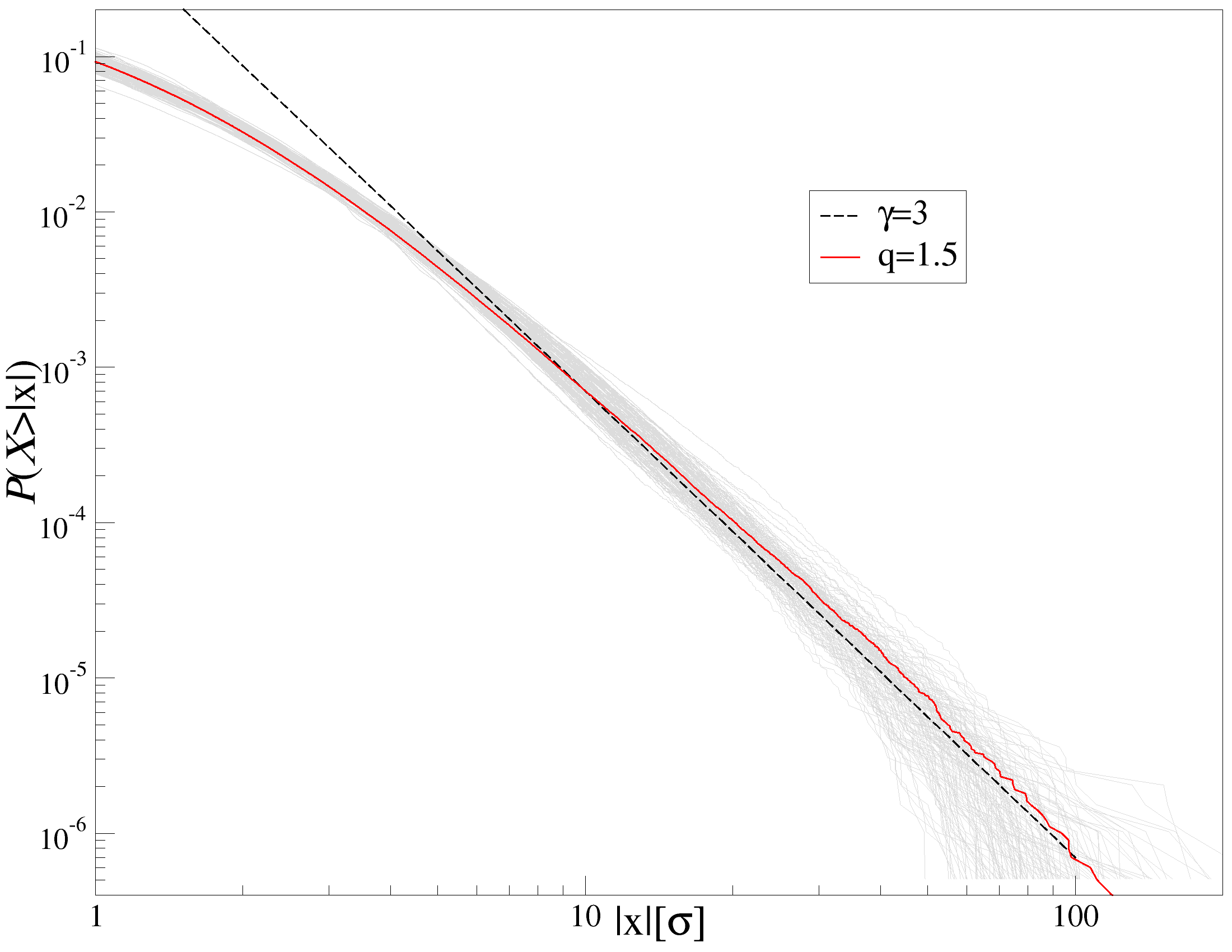}
\caption{Complementary cumulative distribution function (CCDF) of the log-returns $R_i(t_k)$ for all 140 cryptocurrencies together with $q$Gaussian ($q=3/2$) and power law ($\gamma=3$) CCDFs.}
\label{fig::return.pdf}
\end{figure}

From the above rates of return their fluctuation distributions can be directly generated. In the complementary cumulative distribution function (CCDF) representation they are displayed in Fig.~\ref{fig::return.pdf}. As can be seen from the also shown comparison to the corresponding theoretical references, a very good fit of this distribution is obtained in terms of the $q$Gaussian with $q=3/2$ which not only provides an overall good fit but also asymptotically corresponds to the inverse cubic law.

The richness of the dynamics of the set of 140 most actively traded cryptocurrencies considered here is illustrated in Fig. 3, where using a moving 7-day window (with a step of one day), the evolution of the largest eigenvalue $\lambda_1$ of the corresponding matrix $\boldsymbol{\rho}_r(s)$ for $r=2$ and $r=4$ and three different values of the detrending scale ($s$=10, 60, and 360 min) is shown. As can be seen, $\lambda_1$ strongly depends on the window location and, in some places, exhausts a large part of the maximum possible value equal to the trace of the matrix which in this case equals 140. In the vast majority of cases, $\lambda_1$ is larger for $r=2$ case matrix than for $r=4$, which means that cross-correlations occur here rather at the level of fluctuations with medium amplitude. However, it happens, especially in the later part of the time period considered here, that $\lambda_1$ for $r=4$ takes higher values, which signals the dominance of cross-correlation at the level of larger fluctuation amplitudes. A more detailed inspection indicates that this is associated with more rapid price changes in the cryptocurrency market at such moments. It should also be noted that on average the value of $\lambda_1$ increases for increasingly larger detrending scales which is an effect reminiscent of the Epps effect~\cite{EppsTW-1979,KwapienJ-2004,Watorek2019}.


\begin{figure}[ht!]
\includegraphics[width=0.99\textwidth]{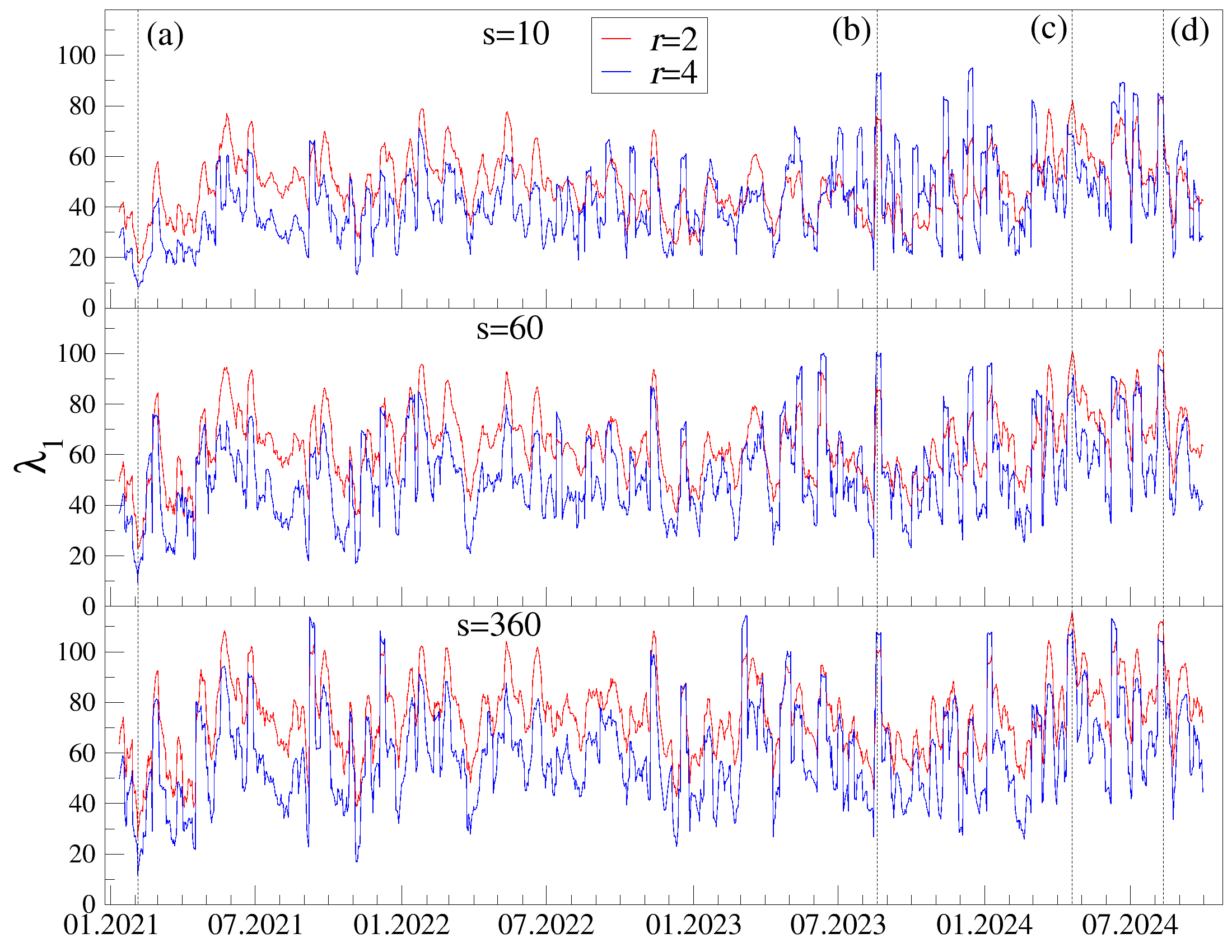}
\caption{Time evolution of the largest eigenvalue $\lambda_1$ of the matrix $\boldsymbol{\rho}_r(s)$ with $r=2$ (red) and $r=4$ (blue) for $s=10$ (top), $s=60$ (middle), and $s=360$ (bottom). A rolling window of length 7 days, shifted by 1 day, was applied. Four specific periods are marked by vertical dashed lines (a)-(d).}
\label{fig::collective.mode.evolution}
\end{figure}

\section{Detrended correlation matrices from random series}

Even for random series, the form of the cross-correlation matrix depends significantly on the parameter $r$, on the detrending scale $s$, and on the fluctuation distributions of these series. As representative examples the synthetic 140 random series of length 10,080 data points (the same as the length of the empirical series) generated from $q$Gaussian distributions, for three values $q=1$, which corresponds to the normal distribution, $q=3/2$, which corresponds to the inverse-cubic power law satisfied by the empirical data considered, and for $q=2$, which corresponds to fluctuations already in the L\'evy-stable regime are used here. 

\subsection{Distribution of matrix elements}

\begin{figure}[ht!]
\includegraphics[width=0.99\textwidth]{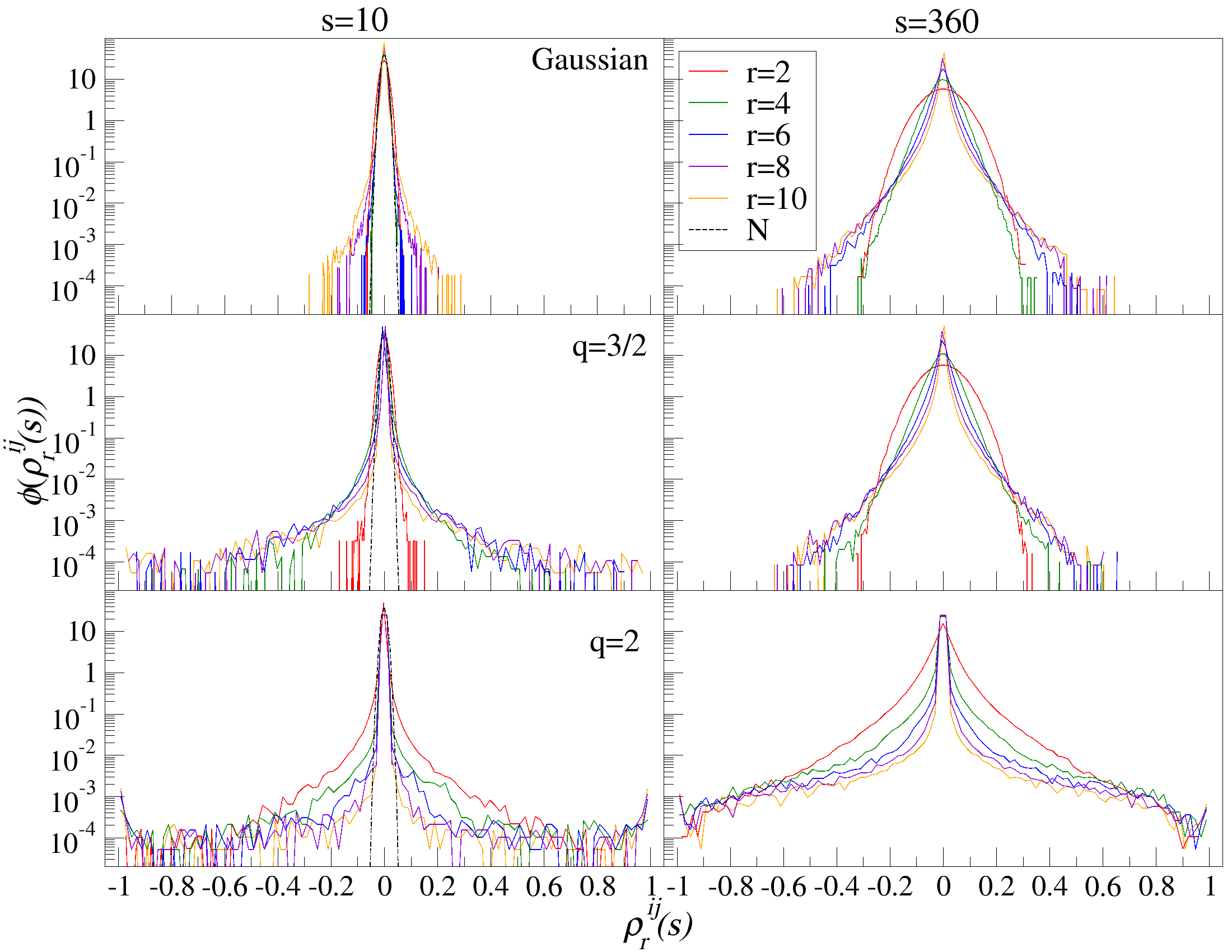}
\caption{Probability density function (PDF) of the off-diagonal elements of the matrix $\boldsymbol{\rho}_r(s)$ obtained from random uncorrelated time series with $q$Gaussian distribution defined by $q=1$ (i.e., a Gaussian distribution, top), $q=3/2$ (middle), and $q=2$ (bottom). Results for a few even values of the index $r$ in a range $2 \leqslant r \leqslant 10$ and two scales $s=10$ (left) and $s=360$ (right) are shown. Each PDF was created from 100 independent realizations of the random data set. The dashed line represents a Gaussian PDF corresponding to random Wishart matrices.}
\label{fig::off-diagonal.pdf.r}
\end{figure}

For these three values of $q$, two significantly different detrending scales $s$ and five values of parameter $r$ from 2 to 10 with step 2, the distributions of off-diagonal entries (diagonals are units by construction) of the corresponding detrended correlation matrix are systematically illustrated in Fig.~\ref{fig::off-diagonal.pdf.r}. As can be seen, all these elements are important and the normal distribution of these entries — as for traditional random matrices — is obtained only for $q=1$, $r=2$ at small detrending scales $s$. Otherwise, the probability distributions of these entries develop increasingly fatter tails as $q$, $r$, and $s$ increase, although this increase is progressing in a slightly different form for each of these parameters. 

\begin{figure}[ht!]
\includegraphics[width=0.99\textwidth]{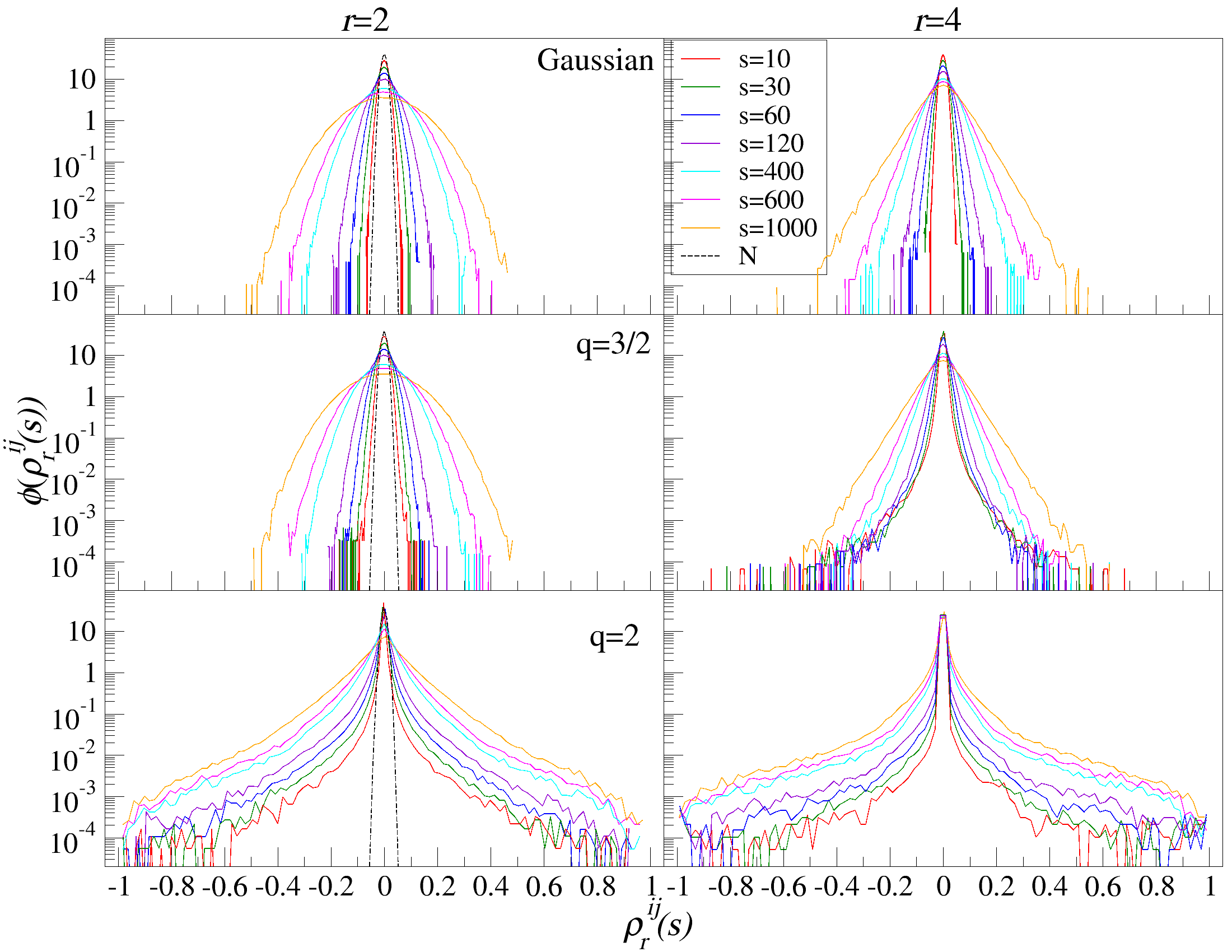}
\caption{Probability density function (PDF) of the off-diagonal elements of the matrix $\boldsymbol{\rho}_r(s)$ obtained from random uncorrelated time series with $q$Gaussian distribution defined by $q=1$ (i.e., a Gaussian distribution, top), $q=3/2$ (middle), and $q=2$ (bottom). Results for different scales $10 \leqslant s \leqslant 1000$ and for two even values of the index $r=2$ (left) and $r=4$ (right) are shown. Each PDF was created from 100 independent realizations of the random data set. The dashed line represents a Gaussian PDF corresponding to random Wishart matrices.}
\label{fig::off-diagonal.pdf.scale}
\end{figure}

Another perspective on these relationships, somewhat complementary, is presented in Fig.~\ref{fig::off-diagonal.pdf.scale}, where more variants of the scale parameter $s$ are considered. A systematic increase in the thickness of the tails of these off-diagonal entries is visible. In general, such effects signal a reduction in the effective rank of the matrix~\cite{DrozdzS-2002}, i.e., the appearance of several large eigenvalues relative to the global bulk, where the dominant portion of small eigenvalues is located.


\subsection{Distribution of eigenvalues}


\begin{figure}[ht!]
\includegraphics[width=0.99\textwidth]{Figs/zaleznoscodq_140_10k_all.png}
\caption{Eigenvalue distribution $\phi(\lambda)$ for the matrix $\boldsymbol{\rho}_r(s)$ obtained from random uncorrelated time series with $q$Gaussian distribution defined by $q=1$ (i.e., a Gaussian distribution, top), $q=3/2$ (middle), and $q=2$ (bottom). Results for a few even values of the index $r$ in a range $2 \leqslant r \leqslant 10$ two scales $s=10$ (left) and $s=360$ (right) are shown. Each PDF was created from 100 independent realizations of the random data set. Dashed black line - corresponding Marčenko-Pastur distribution.}
\label{fig::eigenvalue.pdf.r}
\end{figure}


\begin{figure}[ht!]
\includegraphics[width=0.99\textwidth]{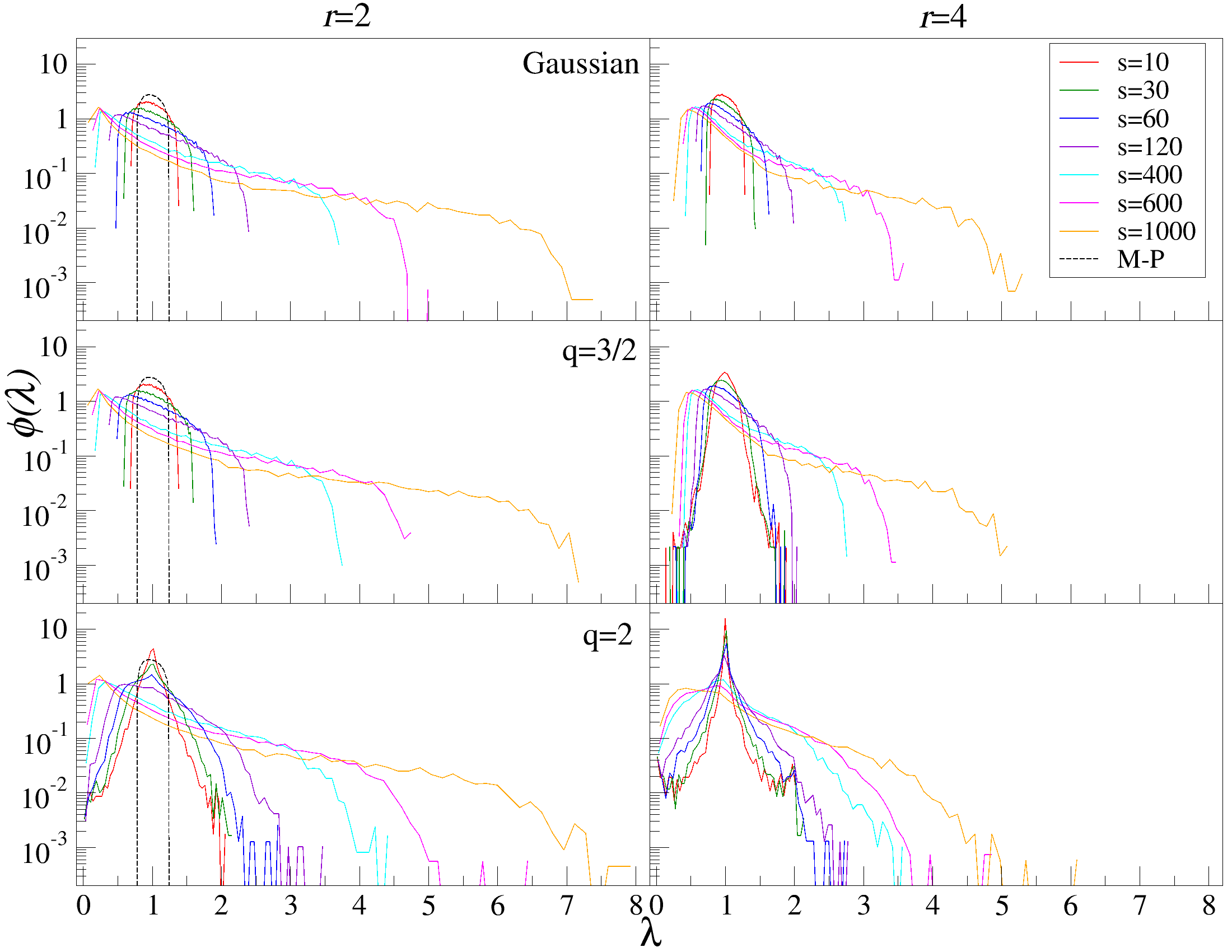}
\caption{Eigenvalue distribution $\phi(\lambda)$ for the matrix $\boldsymbol{\rho}_r(s)$ obtained from random uncorrelated time series with $q$Gaussian distribution defined by $q=1$ (i.e., a Gaussian distribution, top), $q=3/2$ (middle), and $q=2$ (bottom). Results for different scales $10 \leqslant s \leqslant 1000$ and for two even values of the index $r=2$ (left) and $r=4$ (right) are shown. Dashed black line - corresponding Marčenko-Pastur distribution.}
\label{fig::eigenvalue.pdf.scale}
\end{figure}

What is typically analyzed first in the context of cross-correlation is the eigenvalue spectrum $\{\lambda_i\}_{i=1}^N$ of the corresponding correlation matrix. For the cases of the matrices in Fig.~\ref{fig::off-diagonal.pdf.r} and Fig.~\ref{fig::off-diagonal.pdf.scale}, the density distributions $\phi(\lambda)$ of the eigenvalues are displayed in Fig.~\ref{fig::eigenvalue.pdf.r} and Fig.~\ref{fig::eigenvalue.pdf.scale}, respectively. It can be clearly seen that, as the deviations of the off-diagonal element distributions from the normal distribution increase, the departures of the eigenvalue distributions from the M-P distribution increase. The detrending scale $s$ turns out to be more effective in this respect, and for larger values of $s$, even the maximum $\phi(\lambda)$ shifts slightly to the left in relation to the M-P distribution. In turn, with a fixed scale, $\phi(\lambda)$ narrows quite significantly towards the M-P distribution with increasing values of the $r$ parameter. Let us recall that all these effects occur already at the level of the set (here 140) of random time series.

\section{Cross-correlations in empirical data}

Potential real correlations in the empirical data are reflected in the deviations of the characteristics of the respective correlation matrices from the results presented above for the matrices corresponding to random time series.
In order to inspect various aspects of such correspondence in varying situations we selected four specific positions of the moving window spanning 7 days, generated the corresponding detrended correlation matrices $\boldsymbol{\rho}_r(s)$ and calculated a complete set of the eigenvalues in each case. The following situations were distinguished: (a) a minimum value of $\lambda_1(r,s,t)$ over the whole period analyzed, which occurred in a window that ends on Feb 4, 2021 (see Fig~\ref{fig::collective.mode.evolution}); the position of this minimum was stable across all scales $s$ and for both considered values of the exponent $r$; (b) the large-amplitude cross-correlations are significantly stronger than the average ones: $\lambda_1(r=4)>\lambda_1(r=2)$ for all considered scales $s$; a sample window with such a property ends on Aug 18, 2023; (c) the opposite case of $\lambda_1(r=2)>\lambda_1(r=4)$; a sample window with such a property ends on Apr 20, 2024; (d) a sample case of an approximate equality $\lambda_1(r=2) \approx \lambda_1(r=4)$, which occurred in a window that ends on Aug 12, 2024.
As in the case of the numerical simulations, we consider two even values of the exponent: $r=2$ and $r=4$, as well as three scales: $s=10$ min, $s=60$ min, and $s=360$ min. 

\subsection{Distribution of matrix elements}


\begin{figure}[ht!]
\includegraphics[width=0.5\textwidth]{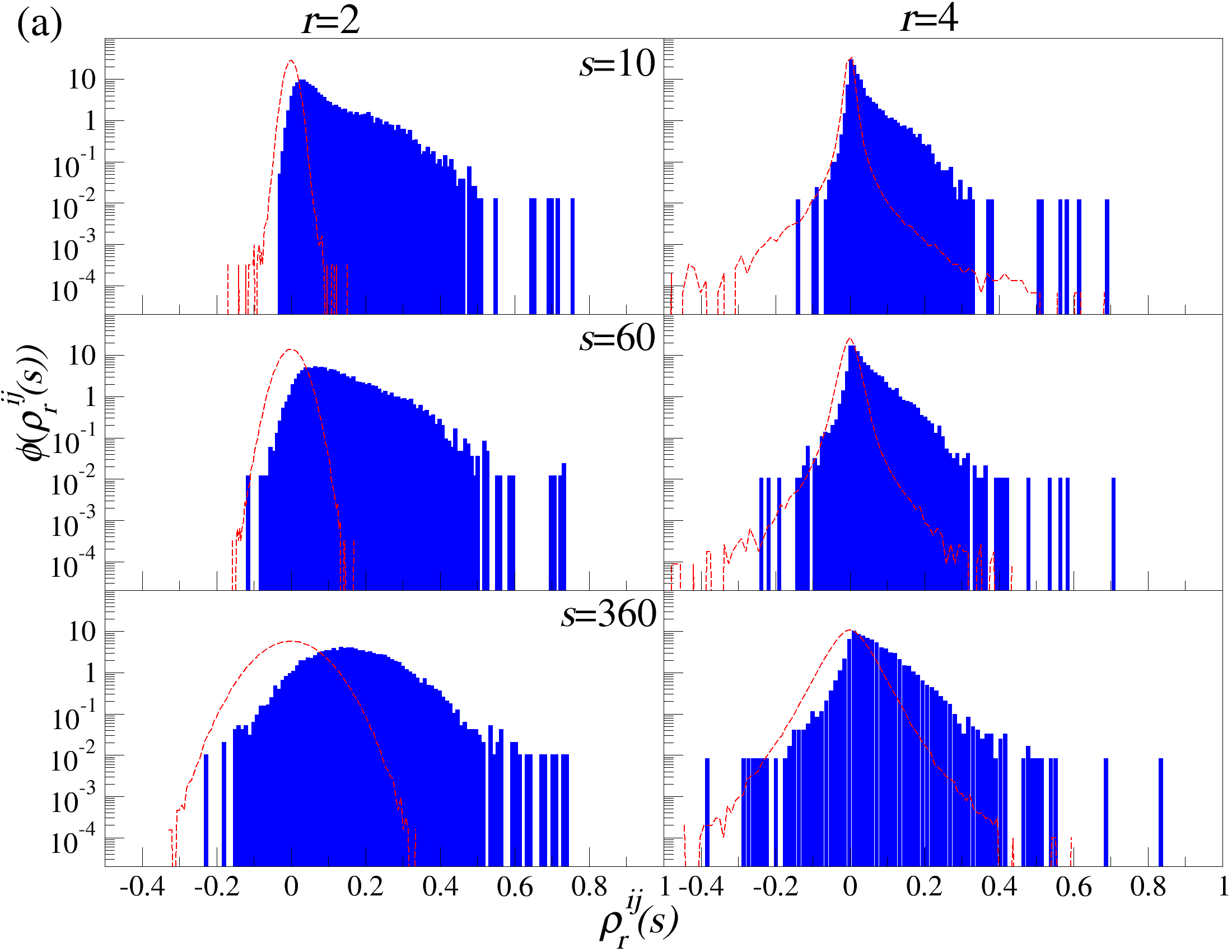}
\includegraphics[width=0.5\textwidth]{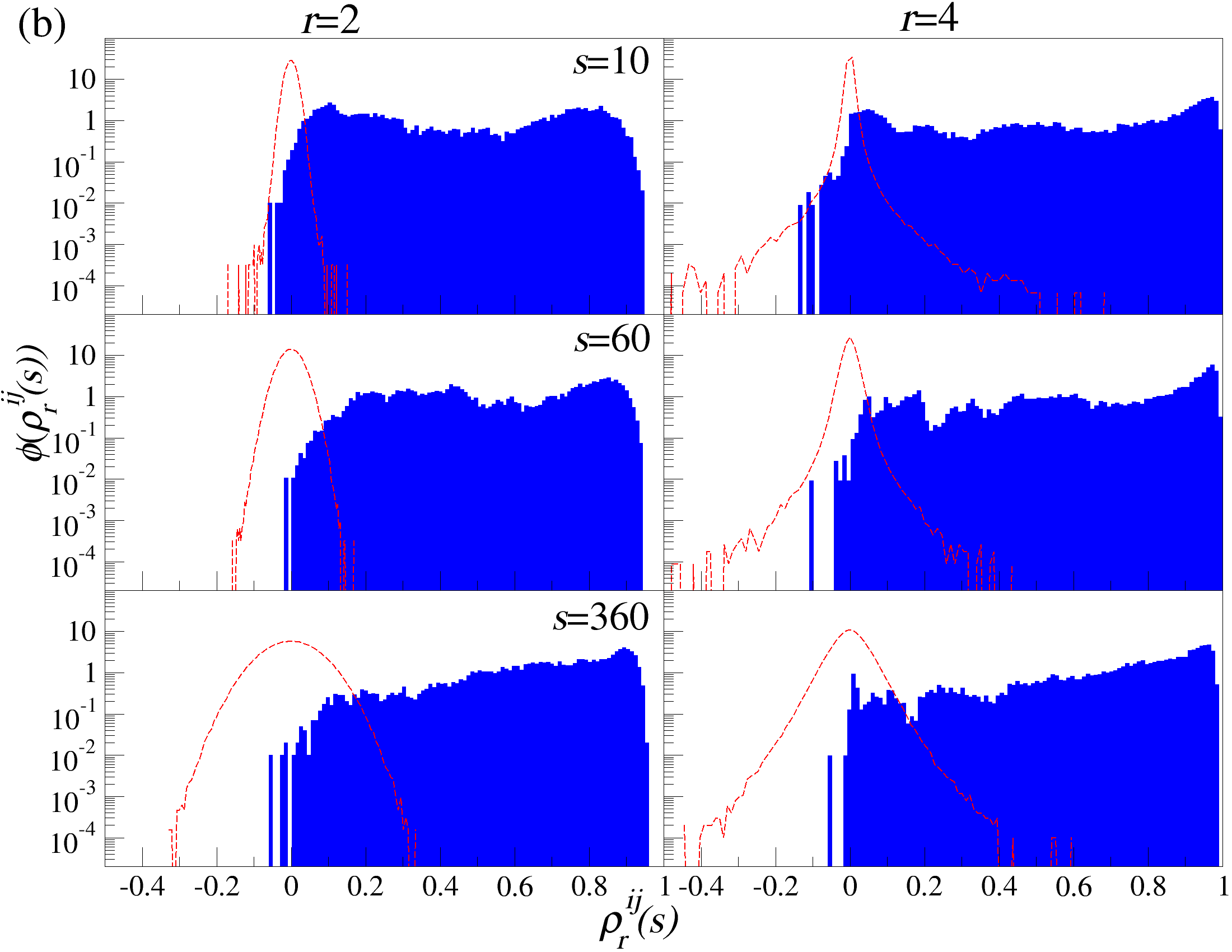}
\includegraphics[width=0.5\textwidth]{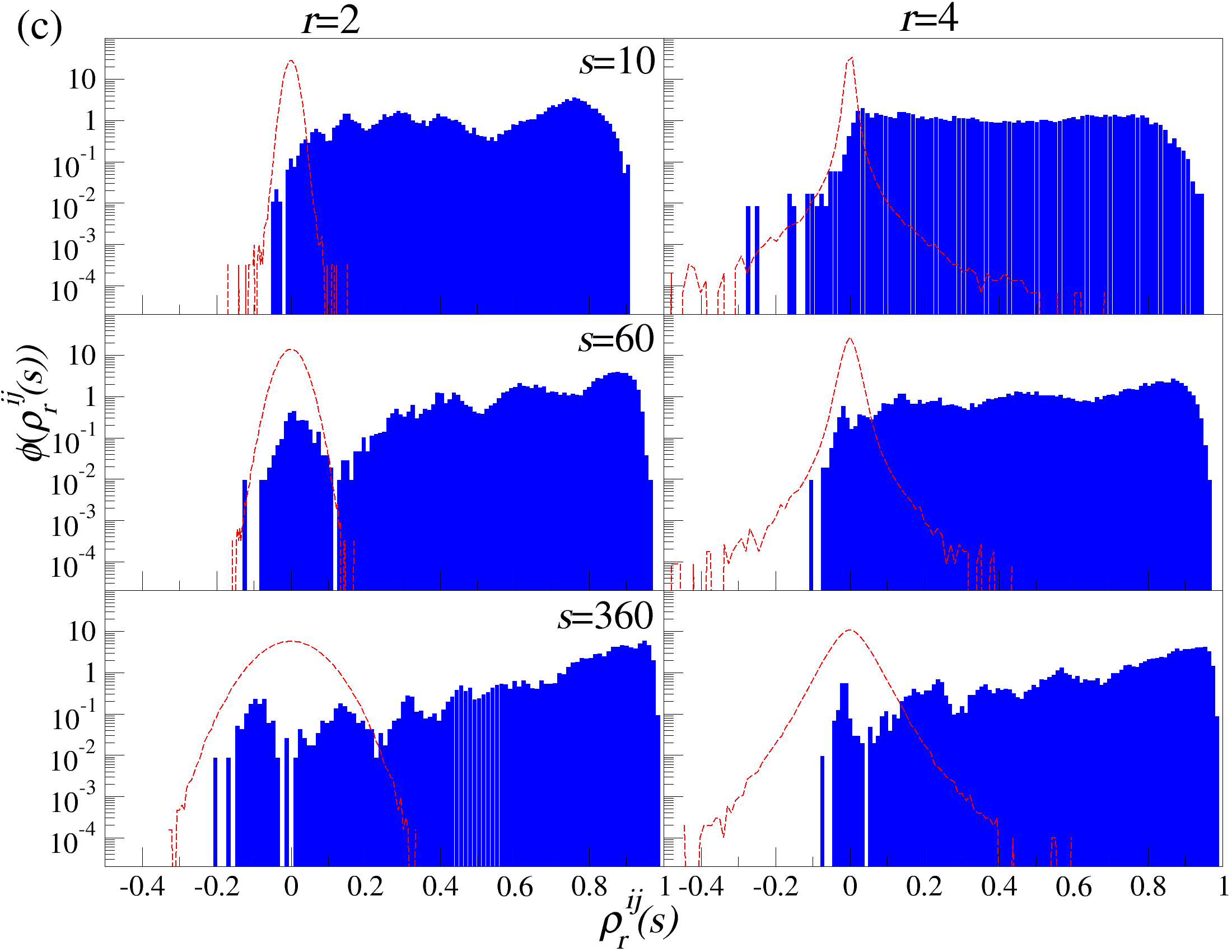}
\includegraphics[width=0.5\textwidth]{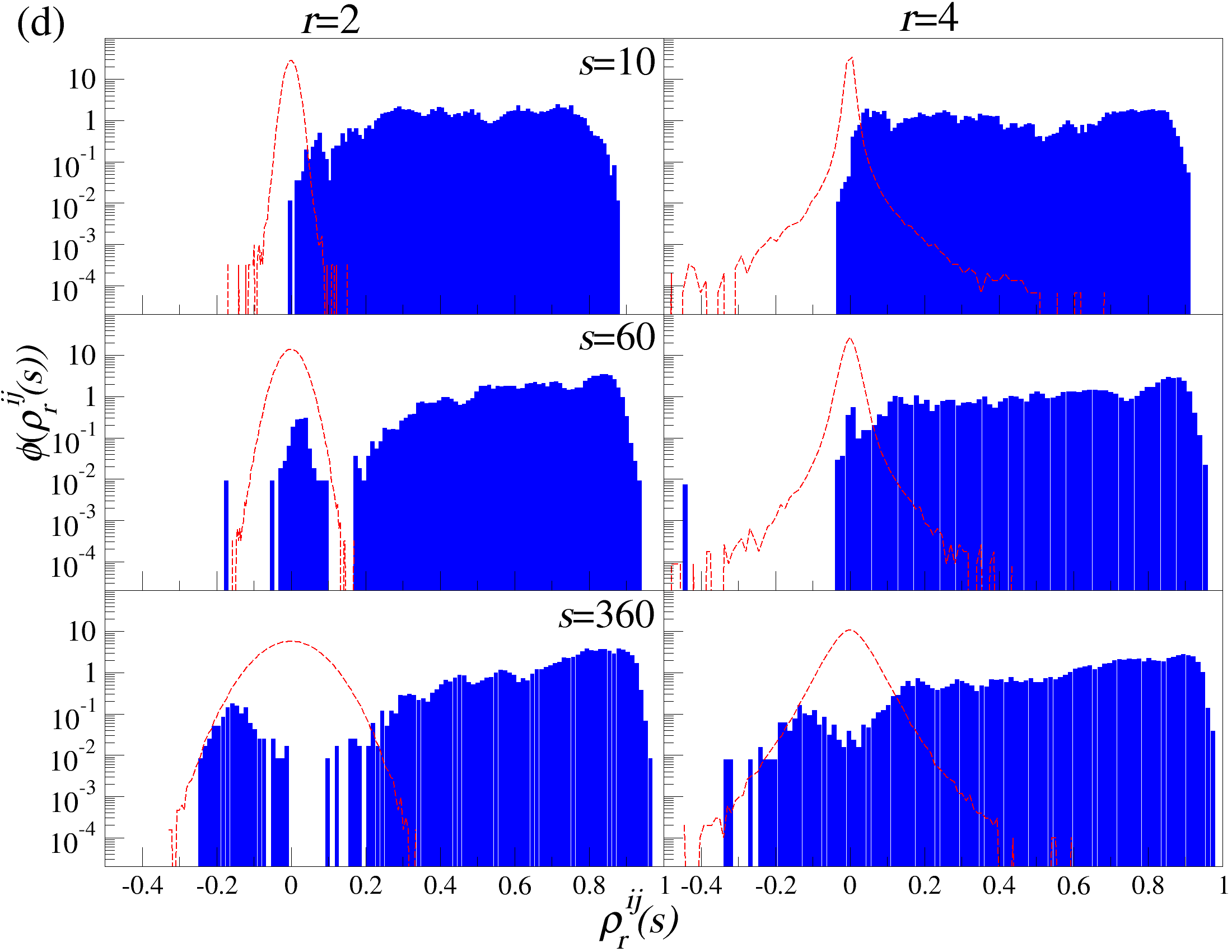}
\caption{Probability density function of the off-diagonal elements of the empirical matrix $\boldsymbol{\rho}_r(s)$ (histogram) calculated in rolling windows ending on (a) Feb 4, 2021, (b) Aug 18, 2023, (c) Apr 20, 2024, and (d) Aug 12, 2024. In each case, results for three scales: $s=10$ (top), $s=60$ (middle), and $s=360$ (bottom) and for two index values: $r=2$ (left) and $r=4$ (right) are shown (histogram) and the corresponding random matrix ensemble (red dashed line). The random case represents a time series with $q$Gaussian distribution with $q=3/2$ and $s$ and $r$ parameters corresponding to cases shown in the middle row of Fig.~\ref{fig::off-diagonal.pdf.scale}.}
\label{fig::off-diagonal.pdf.empirical}
\end{figure}

The distributions of the off-diagonal matrix elements of the resulting correlation matrices are displayed in Fig.~\ref{fig::off-diagonal.pdf.empirical} where their random counterparts for $q=3/2$ (inverse-cubic power law) are indicated by the red dashed lines. The deviations of the empirical results from the corresponding random ones are very pronounced (and the weakest for case (a)) and consist mainly in the appearance of a large number of sizable matrix elements, many of them even close to unity. These effects signal an effective reduction in the dimension of the leading component of the correlation matrix, i.e., the appearance of one large or at most several larger eigenvalues repelled from many small ones, which are remnants of the random part of this matrix~\cite{DrozdzS-2002}.

\subsection{Distribution of eigenvalues}


\begin{figure}[ht!]
\includegraphics[width=0.5\textwidth]{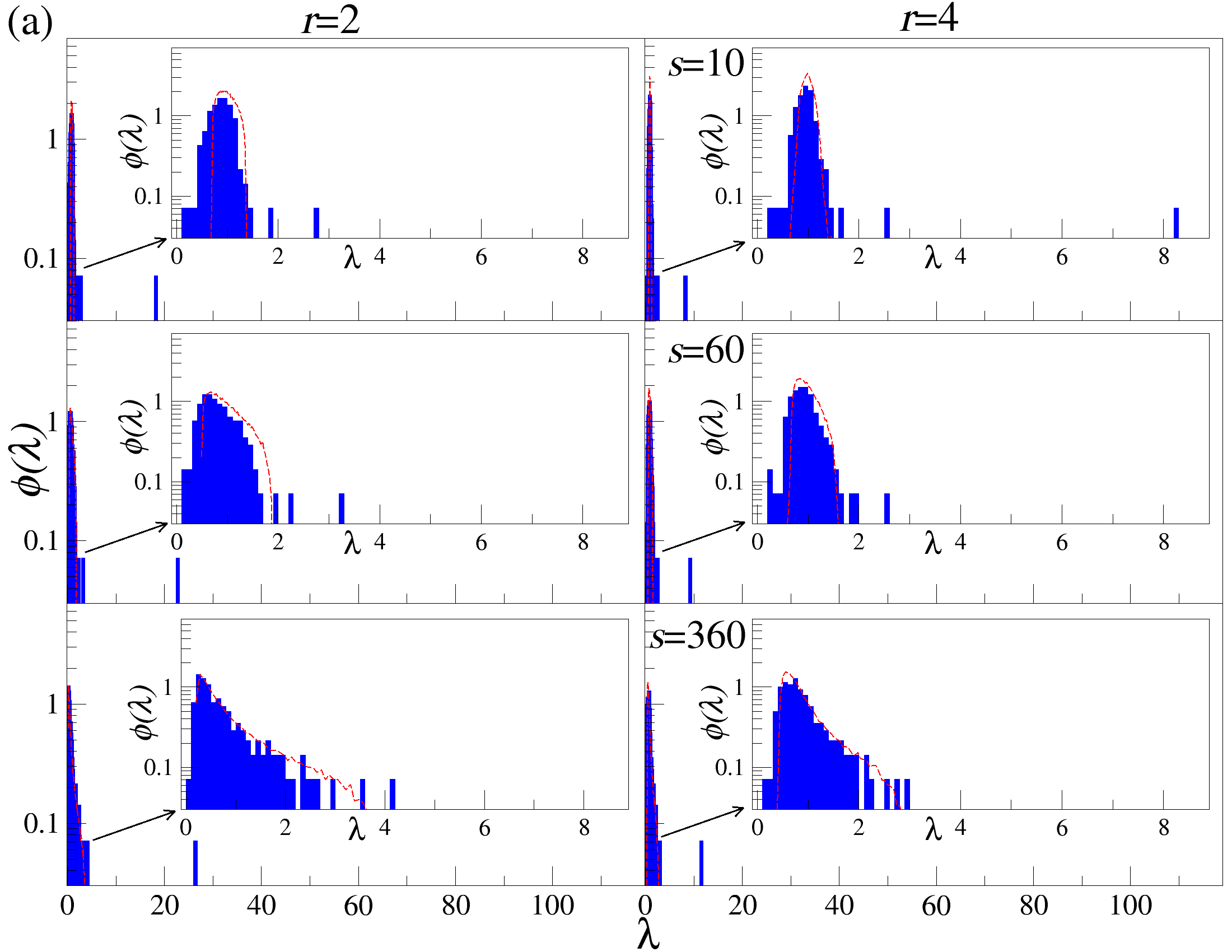}
\includegraphics[width=0.5\textwidth]{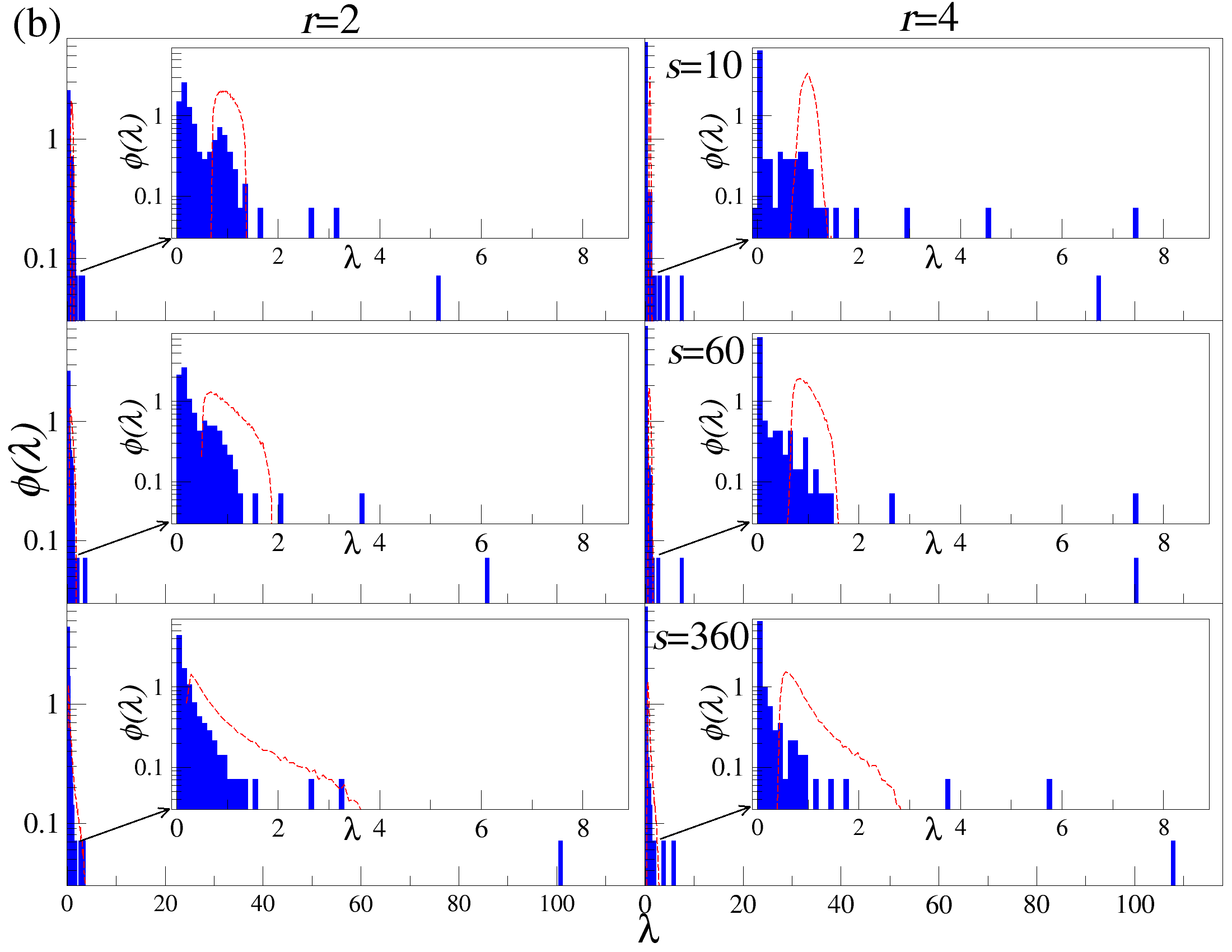}
\includegraphics[width=0.5\textwidth]{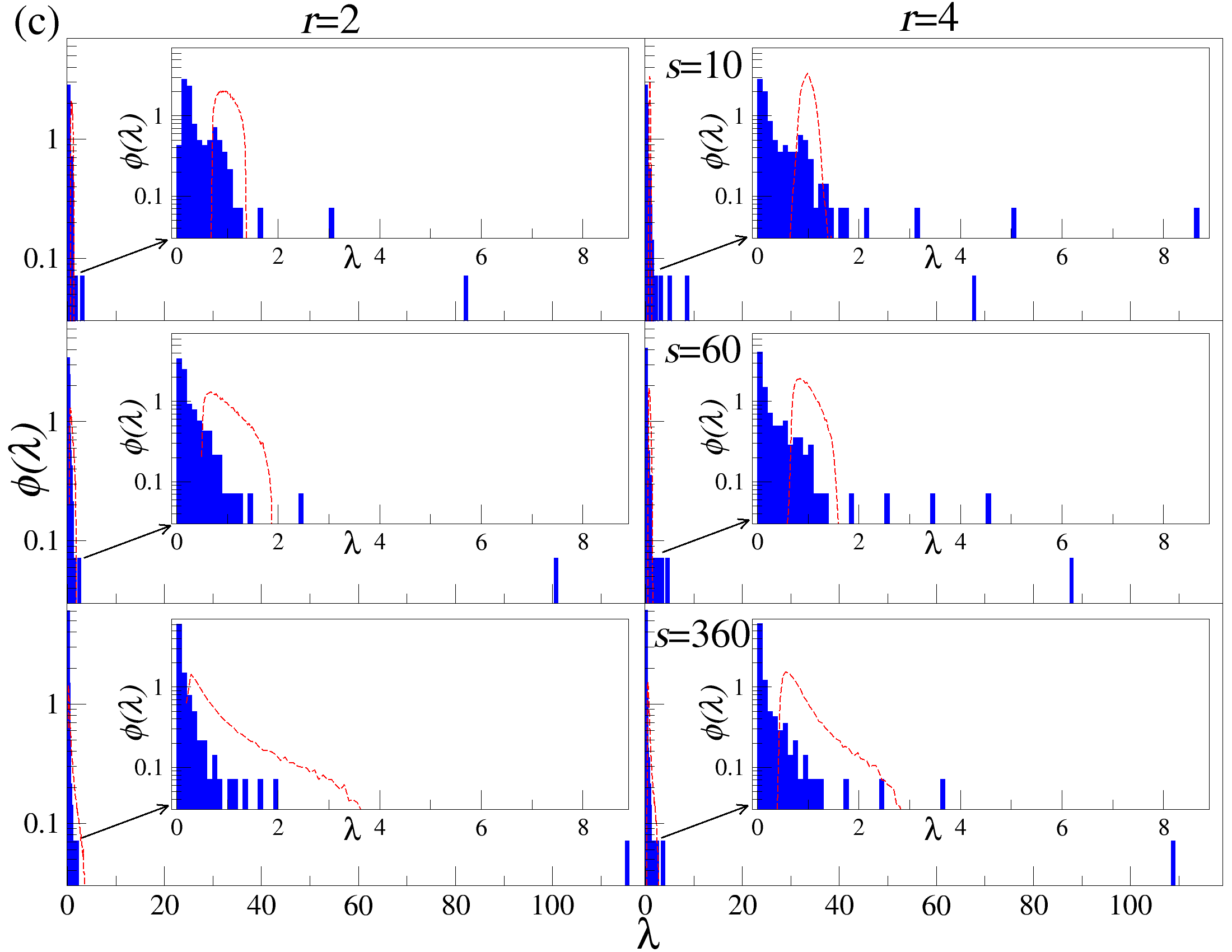}
\includegraphics[width=0.5\textwidth]{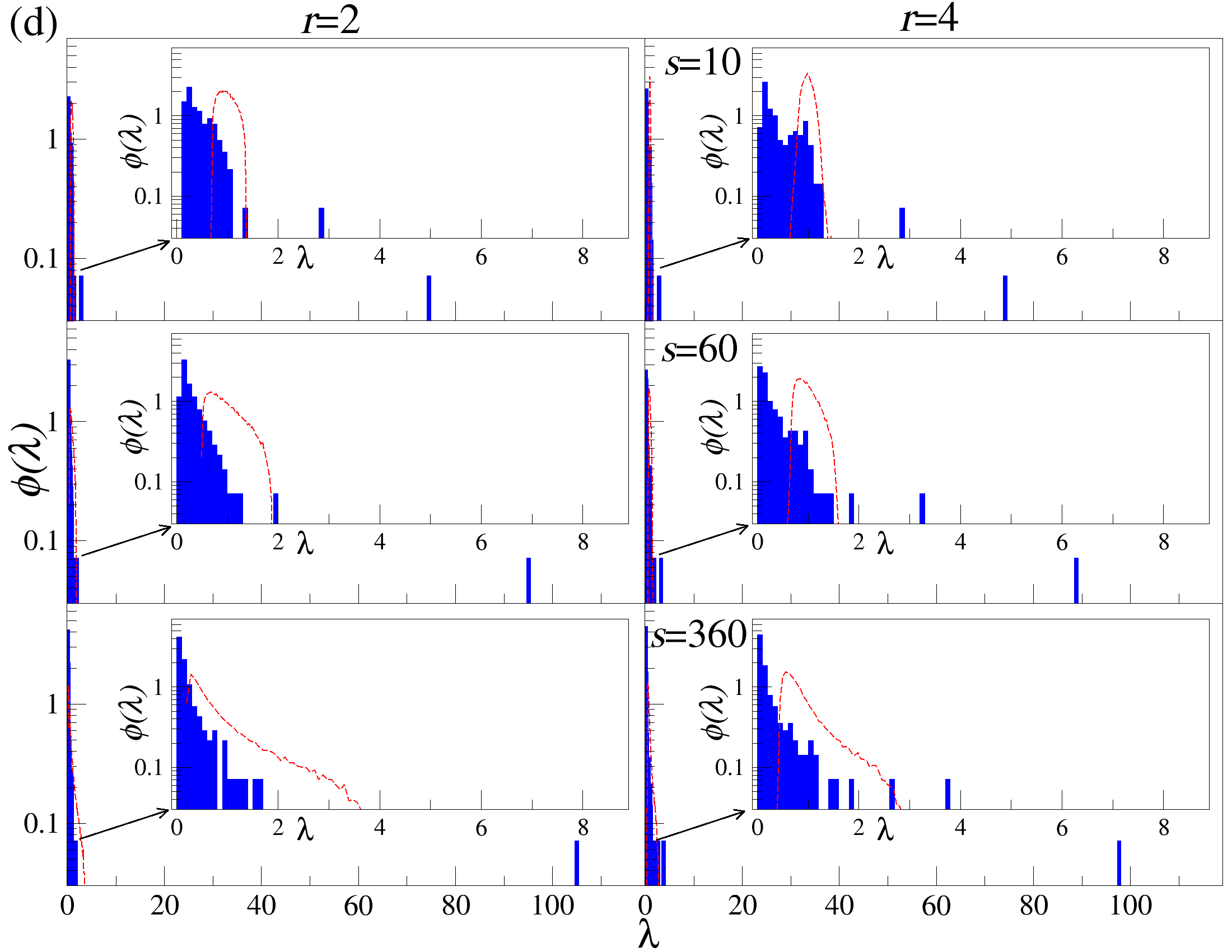}
\caption{(Main) Eigenvalue distribution $\phi(\lambda)$ for the matrix $\boldsymbol{\rho}_r(s)$ obtained from the empirical data in rolling windows ending on (a) Feb 4, 2021, (b) Aug 18, 2023, (c) Apr 20, 2024, and (d) Aug 12, 2024. In each case, results for three scales: $s=10$ (top), $s=60$ (middle), and $s=360$ (bottom) and for two index values: $r=2$ (left) and $r=4$ (right) are shown (histogram). Eigenvalue distributions representing random data with $q$Gaussian distribution with $q=3/2$ and $s$ and $r$ parameters corresponding to cases shown in middle row in Fig.~\ref{fig::eigenvalue.pdf.r} are denoted by red dashed line in each panel. (Inset) Magnification of the bulk-$\lambda$ region of the main panel.}
\label{fig::eigenvalue.pdf.empirical}
\end{figure}

The resulting eigenvalue probability density functions for these four cases denoted by (a)-(d) are shown in Fig.~\ref{fig::eigenvalue.pdf.empirical}. These eigenvalue spectra are compared with the corresponding eigenvalue distribution obtained for uncorrelated time series with an inverse cubic power-law pdf ($q=3/2$, see Fig.~\ref{fig::eigenvalue.pdf.scale}). The same pattern is always observed: the largest eigenvalue $\lambda_1$ is separated from the rest, and the longer the scale $s$ under analysis is, the larger the gap between $\lambda_1$ and $\lambda_2$ becomes. This result is characteristic of all financial markets, where $\lambda_1$ can be associated with the collective movement of prices of all assets (the market factor) and the strength of this collectivity increases as longer return intervals $\Delta t = t_{k+1}-t_k$ are considered. This comes as a natural consequence of delays in the information transfer between the assets stemming from the fact that transactions on different assets are asynchronous and take place at random moments. The market factor is not a unique collective aspect of the cross-correlations among cryptocurrencies: depending on the position of the rolling window on the time axis and on the values of the parameters $r$ and $s$, a 2nd one or more eigenvalues appear to be larger than it would be expected from the eigenvalue distribution for random data. Such individual detached eigenvalues typically describe the sectoral structure of the market, in which groups of assets are more strongly correlated within the group than outside it. Depending on the specific case, in Fig.~\ref{fig::eigenvalue.pdf.empirical} the number of such eigenvalues ranges from 1 to 5, with the typical pattern being that for $r=4$ there are more of them than for $r=2$. It is worth noting, however, that such eigenvalues lying outside the range predicted for random data also appear below the lower bound $\lambda_{\rm min}$ of the random spectrum. In that case, such small eigenvalues describe correlations between individual pairs of assets, without any signs of broader collectivity.

\subsection{Filtering out the market factor}

\begin{figure}[ht!]
\includegraphics[width=0.5\textwidth]{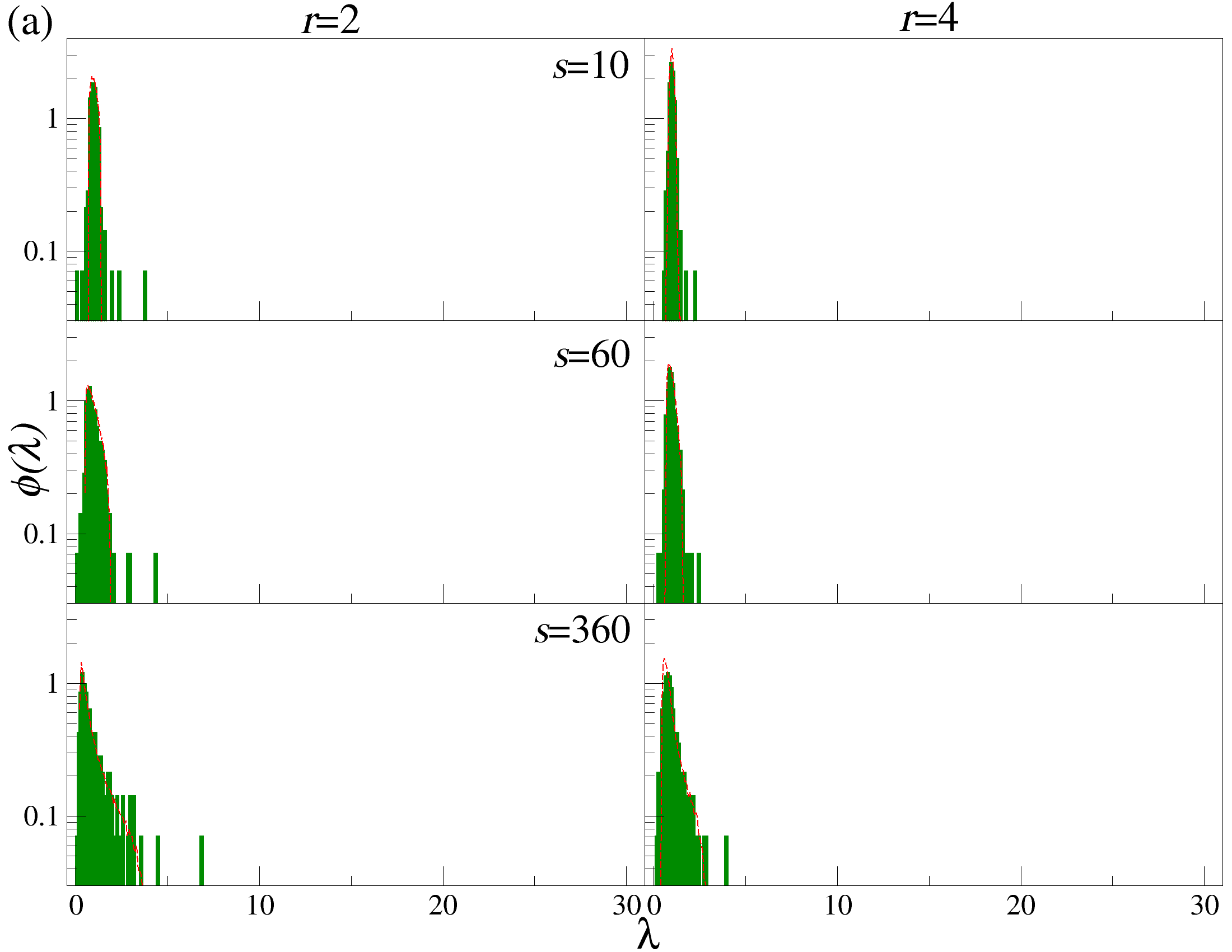}
\includegraphics[width=0.5\textwidth]{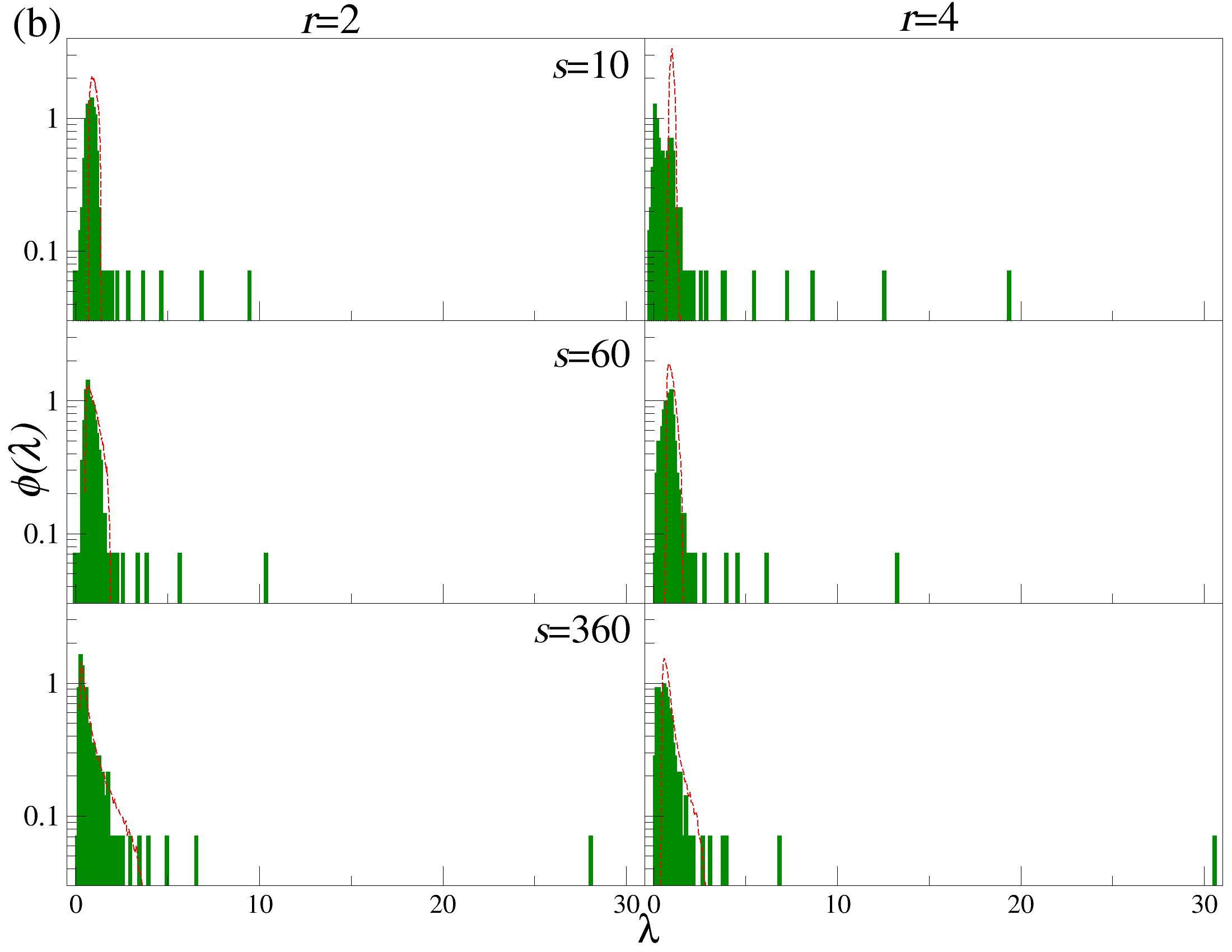}
\includegraphics[width=0.5\textwidth]{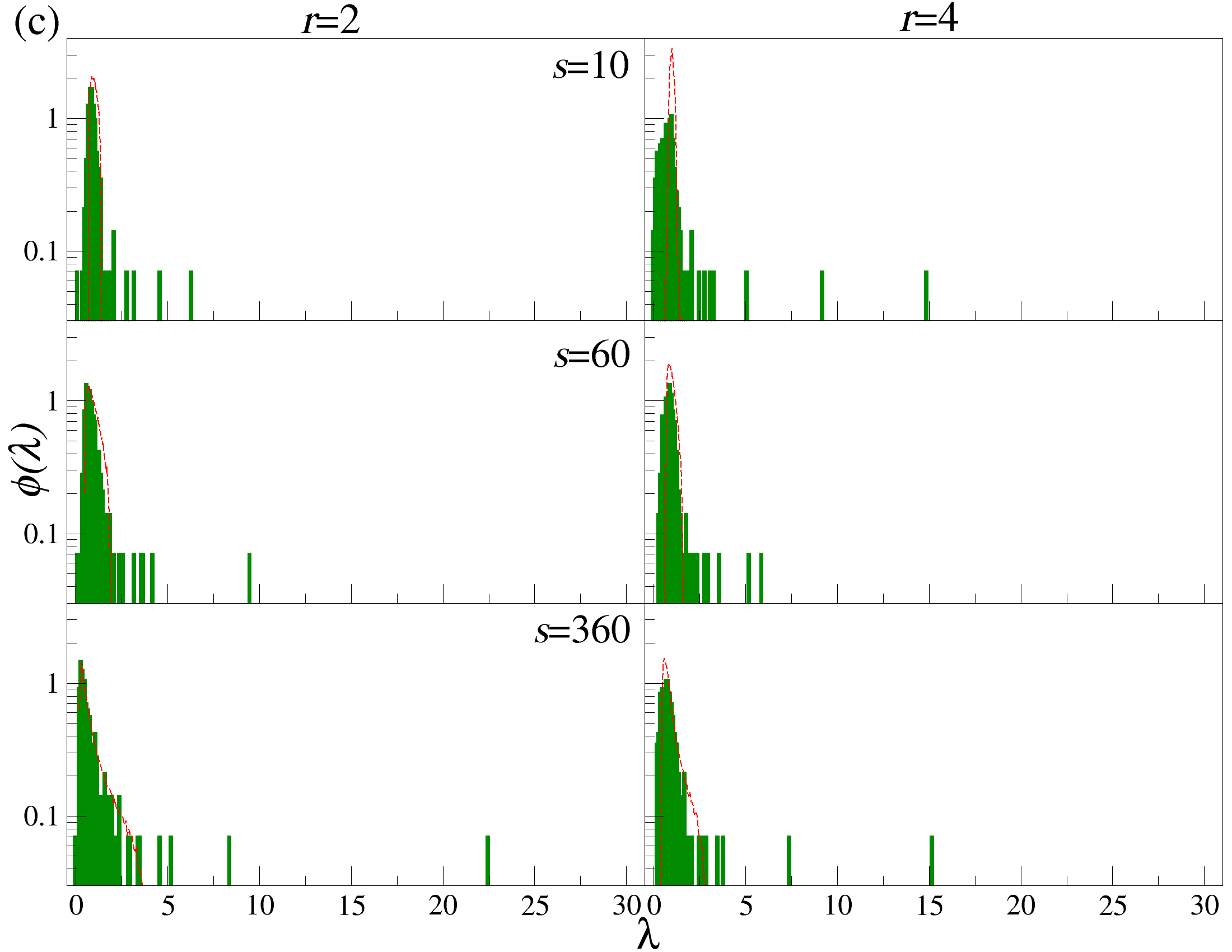}
\includegraphics[width=0.5\textwidth]{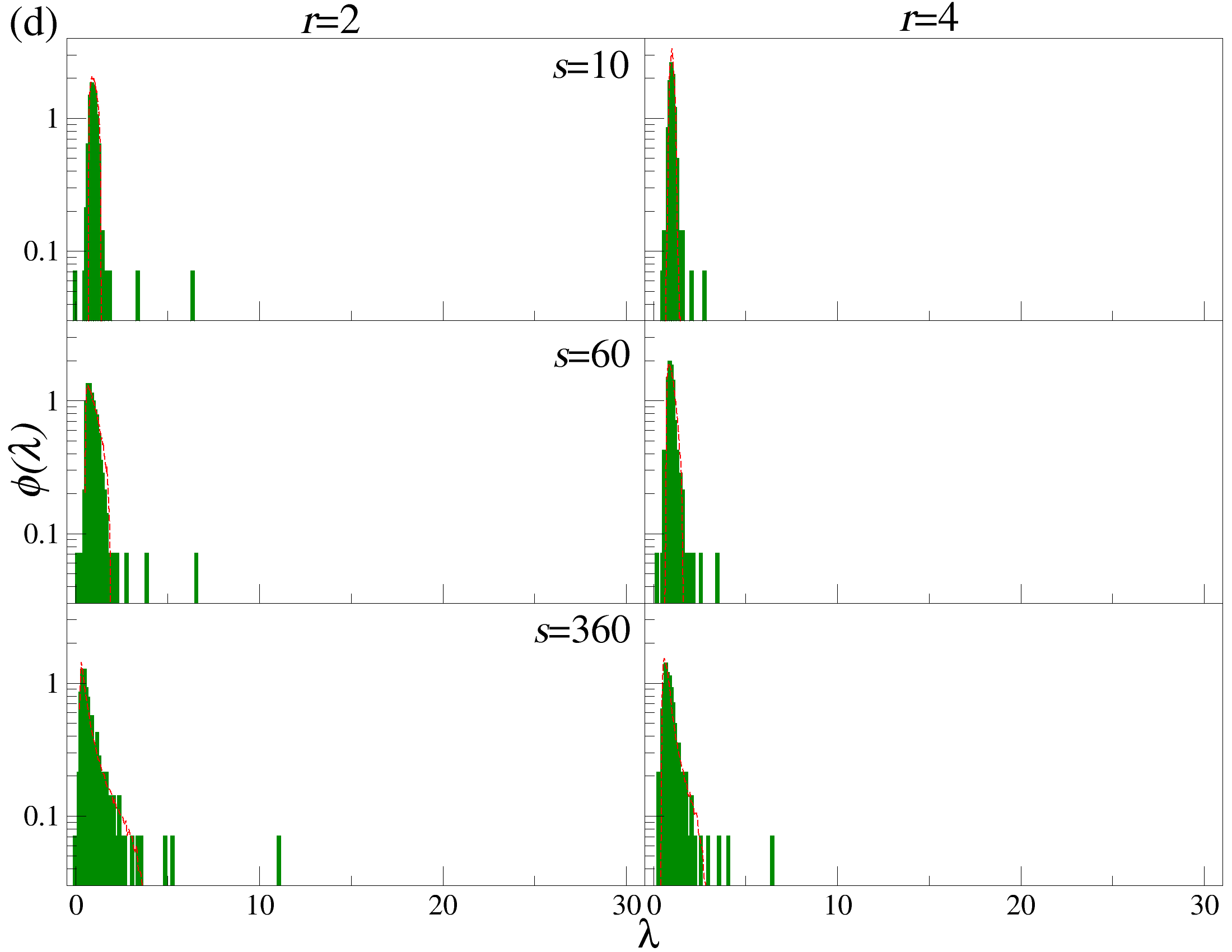}
\caption{Eigenvalue distribution $\phi(\lambda)$ for the filtered matrix $\boldsymbol{\rho}_r'(s)$ obtained from the empirical data in rolling windows ending on (a) Feb 4, 2021, (b) Aug 18, 2023, (c) Apr 20, 2024, and (d) Aug 12, 2024. In each case, results for three scales: $s=10$ (top), $s=60$ (middle), and $s=360$ (bottom) and for two index values: $r=2$ (left) and $r=4$ (right) are shown (histogram). Eigenvalue distributions representing random data with $q$Gaussian distribution with $q=3/2$ and $s$ and $r$ parameters corresponding to cases shown in middle row in Fig.~\ref{fig::eigenvalue.pdf.r} are denoted by red dashed line in each panel.}
\label{fig::eigenvalue.pdf.empirical_filtered}
\end{figure}

An important feature of the spectra shown in almost all panels of Fig.~\ref{fig::eigenvalue.pdf.empirical} is also a sizable displacement towards zero of the bulk of empirical eigenvalues relative to the corresponding eigenvalue distribution for random data. One of the key causes of this displacement is a fact that when the matrix trace is fixed, the presence of a large, repelled eigenvalue $\lambda_1 \gg \lambda_{\rm max}$ and several other eigenvalues above $\lambda_{\rm max}$ reduces the available range for the remaining eigenvalues, which effectively shifts them leftward (reminiscent of the slaving principle of synergetics~\cite{HakenH-2004}). To reduce this effect, it is necessary to filter out the market factor and construct an analogous eigenvalue distribution for the residual data. This can be done by using a regression-based method~\citep{PlerouV-2002a,KwapienJ-2012a}:
\begin{eqnarray}
\nonumber
R^{\rm res}_i(t_k)=R_i(t_k) - a_i - b_i Z_1(t_k) \\
Z_1(t_k) = \sum_{m=1}^N v_{1m} R_m(t_k),
\label{eq::regression}
\end{eqnarray}
where $Z_1(t_k)$ is the contribution to total variance associated with $\lambda_1$ and the filtered matrix $\boldsymbol{\rho'}_r(s)$ is constructed from the residual time series $R^{\rm res}_i(t_k)$ with $i=1,...,N$ and $k=1,...,T$. The results of diagonalization are analogously illustrated in Fig.~\ref{fig::eigenvalue.pdf.empirical_filtered}. Indeed, the bulk of the eigenvalues of the empirical correlation matrix coincides quite well with that of the matrix generated from random series with the same detrending scales $s$ and other parameters. Some eigenvalues slightly pushed towards larger values remain and they reflect more local and sectoral cross-correlations.


\section{Summary}

This study investigates the spectral properties of detrended cross-correlation matrices constructed from the multifractal detrended cross-correlation coefficient $\rho_r(s)$ which generalizes the Pearson correlation to fluctuation-dependent and scale-dependent settings. Because real world multivariate time series — especially financial ones — exhibit strong nonstationarities, heavy-tailed distributions, and long-range dependence, traditional covariance-based correlation matrices often generate misleading structures. To address this, the paper examines how detrending, fluctuation-selection via the parameter $r$ and empirical heavy-tailed distributions collectively influence the eigenvalue spectra of empirical detrended correlation matrices and their counterparts constructed from random time series. Synthetic ensembles of Gaussian and $q$Gaussian time series are first used to characterize the “null” spectral behavior of detrended correlation matrices. The results show that detrending modifies matrix-element distributions and eigenvalue spectra in systematic ways—most notably by fattening the tails of off-diagonal entries and producing departures from the standard Marčenko–Pastur (M-P) distribution even in fully uncorrelated data. The extent of these deviations grows with the detrending scale $s,$ the fluctuation-order parameter $r$, and the heaviness of the underlying distribution tails.

The empirical analysis employs 1 min returns of 140 the most liquid cryptocurrencies traded on Binance between 2021 and 2024. Across rolling windows and multiple detrending scales, the leading eigenvalues of the detrended correlation matrices consistently rise above their bounds for uncorrelated time series, indicating strong collective behavior. The dominant mode corresponds to market-wide synchronous motion, while several subleading modes reflect sectoral or structural groupings within the cryptocurrency market. The number and size of these outlying eigenvalues depend on both the temporal scale and the chosen fluctuation order with $r=4$ generally highlighting cross-correlations among large-amplitude price movements. By filtering out the dominant market mode and reanalyzing the residual covariance structure, the paper demonstrates that the empirical bulk eigenvalue distribution aligns closely with the corresponding null spectra derived from detrended random series. The remaining outliers then cleanly reveal local, sector-specific dependencies. 

Overall, the work provides a refined spectral benchmark - beyond the classical M-P framework - suitable for systems in which detrending and heavy-tailed fluctuations cannot be neglected. Future research could extend this framework by deriving analytical approximations for the random-matrix limits of detrended correlation spectra, helping to formalize the empirical baselines established in this work.

\vspace{6pt} 

\authorcontributions{Conceptualization, S.D., J.K and M.W.; methodology, S.D., J.K. and M.W.; software, M.W.; validation, S.D., J.P., J.K. and M.W.; formal analysis, S.D., J.K. and M.W.; investigation, S.D., J.P., J.K., M.S. and M.W.; resources, M.S. and M.W.; data curation, M.S. and M.W.; writing---original draft preparation, S.D. and J.K.; writing---review and editing, S.D., J.P., J.K. and M.W.; visualization, M.S. and M.W.; supervision, S.D.; project administration, S.D. J.P., J.K. and M.W.; funding acquisition, J.P. All authors have read and agreed to the published version of the manuscript.}

\funding{This research received no external funding.}

\institutionalreview{Not applicable.}

\informedconsent{Not applicable.}

\dataavailability{The data are available in an open repository~\citep{DataBinance}} 

\conflictsofinterest{The authors declare no conflicts of interest.} 

\reftitle{References}
\bibliography{refs}

\end{document}